%% file: root.tex
\documentclass[letterpaper, 10 pt, conference]{ieeeconf}  

\IEEEoverridecommandlockouts                              

\usepackage{pgfplots}
\usepackage{xcolor}
\usepackage{amsmath}
\pgfplotsset{compat=newest}
\usepgfplotslibrary{groupplots}
\usepackage{tikz}
\usetikzlibrary{shapes.geometric, arrows, positioning,backgrounds,fit}
\usetikzlibrary{external}
\usepgfplotslibrary{dateplot}
\usepackage{multirow}

\title{\LARGE \bf
Benefit evaluation of V2X-enhanced braking in view obstructed crossing use cases
}

\author{Jan Zimmermann$^{1}$, Ignacio Llatser$^{1}$, Michael Scherl$^{2}$, Florian Wildschütte$^{1}$ and Frank Hofmann$^{1}$
\thanks{$^{1}$ Corporate Research - Advanced Solutions for Visual Perception, Positioning and Communication Department, Robert Bosch GmbH, 31139 Hildesheim, Germany.
{\tt\small JanChristian.Zimmermann@de.bosch.com}}
\thanks{$^{2}$ Corporate Research - Advanced Autonomous Systems Department, Robert Bosch GmbH, 70465 Stuttgart, Germany.
}}

\begin{document}

\maketitle
\thispagestyle{empty}
\pagestyle{empty}

\begin{abstract}
If a crash between two vehicles is imminent, an Automatic Emergency Brake (AEB) is activated to avoid or mitigate the accident. However, the trigger mechanism of the AEB relies on the vehicle's onboard sensors, such as radar and cameras, that require a line of sight to detect the crash opponent. If the line of sight is impaired, for example by bad weather or an obstruction, the AEB cannot be activated in time to avoid the crash. To deal with these cases, a 2-stage braking system is proposed, where the first stage consists of a partial brake that is triggered by Vehicle-to-everything (V2X) communication. The second stage is composed of the standard AEB that is triggered exclusively by an onboard sensor detection. The performance of this V2X-enhanced 2-stage braking system  is analysed in obstructed crossing use cases and the results are compared against the use of an AEB-only system. The benefit is quantitatively assessed by determination of the crash avoidance rate and, if the crash cannot be avoided, by estimation of the crash severity mitigation.
\end{abstract}

\section{Introduction}\label{sec:introduction}
Vision zero describes the long-term goal to reduce deaths and serious injuries in traffic accidents to zero. The average human driver is prone to making mistakes and thus creating critical situations. Additionally, human crash avoidance capabilities are limited by anatomic constraints, i.e., reaction time. To reduce the number of deaths and injuries, modern cars support the human driver by diverse assistance functions. One of the most crucial functions for crash avoidance and mitigation is automated braking, in its most critical form denoted as Automated Emergency Brake (AEB). This function activates a full force brake if an imminent crash is detected that cannot be avoided by the driver anymore. 

The triggering of an AEB relies on data readings provided by the vehicle’s sensor system, usually comprised of cameras, radar and possibly lidar. However, these onboard sensing capabilities require a line of sight between the sensor and a critical object. This limitation is also applicable to human perception. The line of sight is often impaired by weather conditions, dense traffic, or physical obstructions in occluded intersections.

Vehicle-to-everything (V2X) communication enables the ego vehicle to exchange information with nearby vehicles, infrastructure, and cloud services. This form of wireless communication requires sending and receiving capabilities as well as the establishment of a secure and reliable connection between the communication partners. However, a line of sight between the communication partners is not necessary. Therefore, V2X can be used as an additional sensor that complements the onboard sensor system in obstructed view scenarios. As a result, automated braking and the execution of crash avoidance measures become possible, even if the opponent is not visible to the driver and the onboard sensor system.

\subsection{State of the art}\label{subsec:stateoftheart}
The benefits of triggering an AEB based on V2X data have been investigated for different use cases, such as cooperative maneuvering at an intersection~\cite{thunberg2021unreliable}, truck platooning~\cite{sidorenko2021safety} and Adaptive Cruise Control (ACC)~\cite{sidorenko2022emergency}. A two-stage hierarchical emergency braking system based on V2X communication~\cite{2021-01-7025} has been shown to fulfill the requirements of several C-NCAP\footnote{The New Car Assessment Program (NCAP) provides consumer information on new cars by performing various driving safety evaluation tests. C-NCAP is the program for China.} test scenarios. The system performs a partial brake followed by a full brake maneuver to avoid a potential collision with a front vehicle, while improving the driver comfort with respect to a standard AEB. 

The fusion of lidar and pedestrian-to-vehicle communications has been used to demonstrate the implementation of an AEB system for platoons of automated shuttles at low speeds~\cite{8392783}. First vehicle trials of an AEB triggered by direct 5G-V2X communications (PC5) have shown a high reliability and lower latency compared to a sensor-based AEB in an idealized scenario~\cite{7974355}.

In all found studies, a full brake maneuver with a deceleration around $9~\mathrm{m/s^2}$ is performed based on V2X data only, which is not feasible with current V2X communication standards, because they do not fulfill the necessary functional safety (ASIL) requirements. 

\subsection{Contributions}\label{subsec:contributions}
Within this work, we present three main contributions:
\begin{enumerate}
    \item 2-stage braking system: Considering the current state of the art regarding safety requirements for active safety systems, we propose the concept of a novel 2-stage braking system that extends current AEB systems by an additional, V2X-enhanced brake as a major contribution. 
    \item Crash severity assessment: To quantify the severity of an occurring crash, a study regarding the correlation between defining crash factors and the probability that the crash participants experience a severe or fatal injury, is conducted. The resulting models are used for the quantification of the V2X benefit.
    \item Benefit evaluation: The benefit of incorporating V2X in an active safety system is evaluated in 35 different scenarios of obstructed crossing use cases, considering bicycle opponents as well as passenger car opponents. The performance of the V2X-enhanced 2-stage brake is compared against an AEB-only solution as a reference regarding its crash avoidance and mitigation capabilities. The results are summarized and discussed in detail.
\end{enumerate}
The paper is structured as follows. In the subsequent Section~\ref{sec:twostagebrakingsystem}, the 2-stage braking scheme is discussed, including a brief overview of V2X communication as a vital part of the system. The following Section~\ref{sec:crashseverityassessment} deals with the crash severity assessment, providing insights in the creation of the crash severity models as well as the presentation of the final models. Before the setup of the simulation and the results of the benefit analysis are presented in Section~\ref{sec:simulation}, we describe in detail the evaluation methodology in Section~\ref{sec:evaluationmethodology}. The latter includes an argumentation for the choice of use cases for the final simulation. The paper closes with a summary and outlook in the Conclusion Section~\ref{sec:conclusion}.

\section{2-stage braking system}\label{sec:twostagebrakingsystem}
We present a novel 2-stage braking system, consisting of a V2X-enhanced first stage partial brake and a sensor-triggered second stage full brake, i.e., an AEB. Before the braking scheme is presented, assumptions regarding the vehicle types as well as the vehicle equipment are made, and a short introduction into V2X communication is provided.

\subsection{Vehicles, equipment and V2X communication}\label{subsec:v2xcommunication}
We focus on crash situations between two vehicles: an \textit{ego vehicle} and an \textit{opponent vehicle}. The ego vehicle aims to avoid or mitigate an imminent crash by use of active safety systems and is therefore equipped with the braking system under test. It is modelled as a passenger car and carries an onboard sensor system, comprised of a combination of video and radar sensors.

The opponent vehicle is the collision adversary of the ego vehicle and is modelled in this work as a passenger car or an (electric) bicycle. It is neither assumed that the opponent vehicle has an onboard sensor system nor that it is equipped with a specific automated braking system. 

Both vehicles are assumed to have V2X communication capabilities and exchange their status via cooperative awareness. Alternatively, the presence of a non-communicating opponent vehicle could be perceived by a V2X-equipped roadside infrastructure and communicated to the ego vehicle via collective perception, meaning that the ego vehicle perceives its environment using data originating from sensors outside of the vehicle.

As described, V2X communication allows nearby vehicles to exchange information with each other, thereby enabling advanced driver assistance and automated driving functions. Currently, there are several competing technologies to support direct V2X communication in the 5.9~GHz frequency band~\cite{anwar2019physical}. IEEE 802.11p, also known as DSRC in USA and ITS-G5 in Europe, is mature and already deployed in Europe. Its backwards-compatible evolution IEEE 802.11bd is currently under development. LTE-V2X and 5G NR-V2X are based on 3GPP cellular technology and have emerged as an alternative to enable direct communication among vehicles using the sidelink or PC5 interface. Cellular-based V2X is already deployed in China and is ready to be deployed in the USA.

Two common V2X applications are \textit{cooperative awareness} and \textit{collective perception}. Cooperative awareness enables vehicles to transmit V2X data regarding their current state, such as their position, speed, and heading. This service has been standardized worldwide, e.g., with the Cooperative Awareness Message (CAM) in Europe and the Basic Safety Message (BSM) in the US. All V2X-equipped vehicles transmit these messages continuously with a variable frequency between 1 and 10 Hz, depending on the variation of their dynamic state, i.e., current heading, position, and speed, as well as the measured channel load.

Collective perception~\cite{gunther2016realizing} allows stations, vehicle and roadside units to inform nearby vehicles of objects, such as pedestrians, obstacles, and other vehicles, detected by their onboard sensors. This enables receiving vehicles to extend their perception capabilities beyond their own sensors’ range. The object data is exchanged through recently standardized Collective Perception Messages (CPM) and Sensor Data Sharing Messages (SDSM) in Europe and the US, respectively.

\subsection{Braking scheme}\label{subsec:twostagebrake}
The first stage of the braking system is denoted as a partial brake. If this brake is triggered, a moderate braking force of 4 $\mathrm{m/s^2}$ is applied with a jerk of 45 $\mathrm{m/s^3}$. Activation occurs based on the opponent vehicle detection by received V2X data and/or by onboard sensors, e.g., through cooperative awareness or collective perception. 

The second stage is an automatic emergency brake (AEB) that aims to decelerate the car with the all available braking force, which is usually estimated as 9 $\mathrm{m/s^2}$ and applied with the same jerk as for the first stage, i.e., 45 $\mathrm{m/s^3}$. In contrast to the first stage, the second stage can only be triggered based on the opponent detection by the vehicle's own onboard sensors. 

The two-staged structure and the hard-coded disregard of V2X information by the second stage are due to the following reason: The information that is used to trigger a brake with  deceleration of more than 4~$\mathrm{m/s^2}$ needs to satisfy a certain \textit{Automotive Safety Integrity Level} (ASIL), if the brake application time should not be limited. In fact, triggering a full force brake such as the AEB requires ASIL B\footnote{ASIL spans from ASIL A, which is the lowest level, to ASIL D, which marks the highest level. Quality management (QM) level ranges below ASIL A and does not require any safety assurance controls~\cite{Gheraibia2018}.}. However, currently, V2X information is not deemed to fulfill ASIL A or higher. Because of that, the braking system is split into the aforementioned two parts: the first stage partial brake that does not require any ASIL due to its limited braking force, therefore allowing the use of V2X data, and the second stage, consisting of the full force AEB, triggered exclusively by the ASIL B compliant onboard sensor system.

Although the chosen labels \textit{first stage} and \textit{second stage} imply that the brakes need to be triggered in succession, situations can arise in which only the first stage or only the second stage is triggered.
\begin{figure}[h]
    \centering
    \input{tikz/twostagebrake_flow}
    \caption{Decision process for 2-stage brake in the ego vehicle.}
    \label{fig:flow_diagram_twostage_brake}
\end{figure}
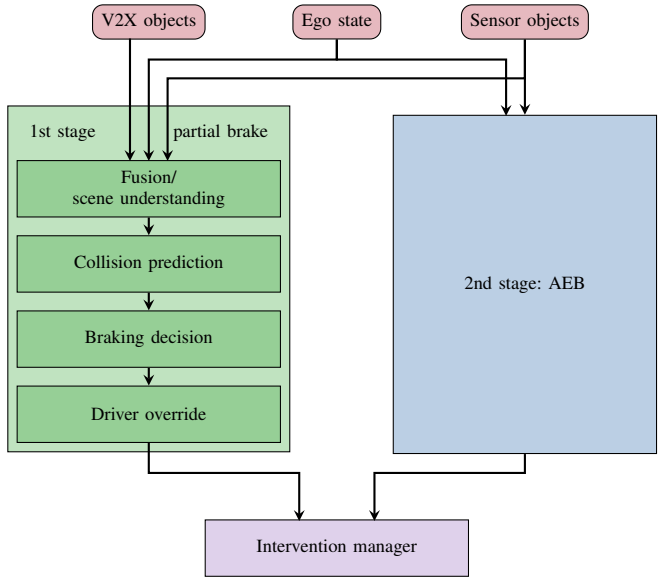
In Figure~\ref{fig:flow_diagram_twostage_brake}, the flow diagram regarding the decision process of the 2-stage brake is depicted. The V2X object interface provides information regarding objects, such as opponents, that are received via V2X communication. The sensor object interface, on the other hand, provides information about detected objects that are recognized by the onboard sensor system.  The ego state provides information such as velocity, position, acceleration and steering angle of the ego vehicle.  

In the first decision block of the partial brake, information from all three interfaces are fused to form a common \textit{scene understanding}, possibly by use of map information as well. In the subsequent \textit{collision prediction} block, the ego vehicle predicts its own path as well as trajectories for all recognized opponent vehicles, using their current perceived status. Based on intersection analysis of the respective trajectories, a potential crash can be detected. In the \textit{braking decision} block, the triggering conditions are evaluated and it is decided if the brake should be triggered in the current time step. As a human driver is in control of the vehicle, they can override the system's braking decision by pressing the gas pedal or performing an evasive steering manoeuvre.

The second stage of the 2-stage braking system is the AEB, which receives information from the ego state interface as well as the sensor objects interface. The decision process regarding this brake is roughly the same as the process of the first stage brake and will not be discussed in detail. 

At last, the intervention manager receives the result from the two brake blocks and decides which brake to trigger. Note that the AEB always has priority over the partial brake. This is due to the fact that the AEB brakes with maximum force and is triggered at the latest possible point in time, as discussed in the subsequent paragraph. Therefore, neither the driver with a slower reaction time nor the partial brake with reduced braking force can reach the same braking performance at this point in time.

\subsubsection{First stage trigger}

When a crash is predicted, two main brake trigger conditions must be fulfilled to activate the first stage brake in the \textit{braking decision} block of Figure~\ref{fig:flow_diagram_twostage_brake}. 

The first trigger condition ensures that the brake is not applied earlier than necessary. To evaluate this condition, the stopping distance $x_\mathrm{stop}$, i.e., the distance that the ego car needs to brake to a full stop, is calculated based on the vehicle's velocity $v$ at the start of the brake application, acceleration $a$, jerk $j$, and brake application delay time $\delta_t$. Assuming a straight path and no driver intervention, the distance can be derived by calculating first the vehicles velocity $v(t)$ under the braking maneuver until it reaches 0, and then integrating over time $t$ to receive the traveled distance, i.e. the stopping distance $x_\mathrm{stop}$. At last, the distance traveled during the brake application delay needs to be considered, which equals $v \delta_t$. Thus, we receive 
\begin{equation}\label{eqn:xstop}
    x_\mathrm{stop} = \frac{1}{2} \frac{av}{j} - \frac{1}{24}\frac{a^3}{j^2} + \frac{1}{2} \frac{v^2}{a} + v \delta_t.
\end{equation}
This distance is then compared against the distance $x_\mathrm{crash}$, which marks the distance that the ego vehicles travels on its predicted trajectory from the current position to the position of the predicted crash with the opponent vehicle. If $x_\mathrm{crash} \leq x_\mathrm{stop}$, the brake should be triggered to avoid or mitigate the imminent crash. 

The second trigger condition prevents the brake from being triggered time-wise too early before the predicted crash. Here, the predicted time to collision (TTC) is compared against some specified upper threshold $\bar{T}$, e.g. $\bar{T}=2s$. The condition is satisfied if $\text{TTC} \leq \bar{T}$. The limitation is necessary, as the earlier the brake is triggered, the higher the possibility that the situation changes even without application of the brake. Due to the limited braking force of the first stage brake in combination with possibly high ego velocities,  the stopping distance  $x_\mathrm{stop}$ grows, requiring an early brake application time. This time is limited by the second condition even if the crash cannot be fully avoided any more. 

In summary, the first stage brake is triggered if both conditions are satisfied, i.e., 
\begin{equation}\label{eqn:conditions}
    x_\mathrm{crash} \leq x_\mathrm{stop}  \text{ and } \mathrm{TTC} \leq \bar{T} .
\end{equation}


\subsubsection{Second stage trigger}
The AEB activation uses the same trigger conditions as the first stage. However, different parameters are used. For the calculation of the stopping distance in Equation~\eqref{eqn:xstop} for example, higher deceleration values are inserted, as the AEB is a full force brake. The second condition in Equation~\eqref{eqn:conditions} plays a more subordinate role compared to the first stage brake, as  the necessary braking time span before the crash is already limited by the shorter stopping distances $x_\mathrm{stop}$ due to higher deceleration values.

\section{Crash severity assessment}\label{sec:crashseverityassessment}
In order to evaluate the performance of a braking system, an assessment method for the crash severity is necessary. In this work, we equate the crash severity with the grade of injury that the direct crash participants experience and disregard other factors such as material damages, secondary collisions and psychological trauma. 

As the estimation of the injury level of arbitrary crash situations is highly complex, we restrict the analysis to situations where an ego vehicle crashes with its front into the side of an opponent vehicle. These front-to-side crashes are most common in obstructed crossing use cases and are as such chosen for the benefit evaluation. For a detailed description of the considered use cases, see the subsequent Section~\ref{sec:evaluationmethodology}. Within this restriction, the injury level is mainly dependent on three factors: the impact velocity, the impact location, i.e., at which position the opponent vehicle is hit by the ego vehicle, and the vehicle type, i.e., passenger car or bicycle.

As a direct calculation of the injury level given these factors is not possible, we analyse the correlation between these factors and the probability of severe/fatal injuries and create respective models based on empirical data. We consider an injury to be severe if the injured person needed to be treated as an inpatient at a hospital. 

\subsection{Models for crash severity probability}\label{subsec:severitymodels}
To analyse the correlation between the impact velocity, impact position, vehicle type and the probability for a severe/fatal injury of the crash participants, the GIDAS\footnote{GIDAS (German In-Depth Accident Study) is a detailed data base, which describes actual crashes that occurred in Germany. Next to many parameters, the injury level of the participants as well as the impact velocity and impact location are saved. However, the data base exclusively contains crashes that resulted at least in a slight injury, while crashes entailing only material damage were not considered.} data base is reviewed. All crashes, in which a passenger car had a front-to-side crash with a passenger car or bicycle, are selected for evaluation. For every crash  $i$, the impact velocity $v_i$ and injury level $s_i$ (coded as a binary variable with 0 for slight injury and 1 for severe/fatal crash) are extracted as data $x_i = (v_i, s_i)$, resulting in the data set $\mathcal{X} = \{x_1, ..., x_i, ..., x_N\}$. This data is used in a logistic regression, where the parameters $a, b$ of the logistic function
$
    f(v_i; a, b) = \frac{1}{1+ \exp(av_i + b)}
$
are chosen optimally by maximum likelihood estimation to fit the function to the data. The function outputs the probability of severe/fatal injury based on the impact velocity as input. Note that due to the absence of crashes without injuries in GIDAS, the probabilities for severe/fatal crashes might be slightly overestimated. 

\begin{figure}
    \centering
    \input{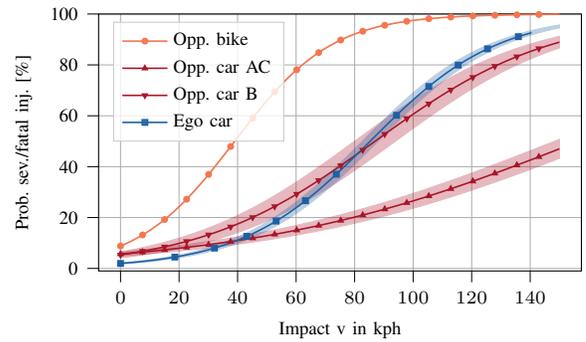}
    \caption{Regression models describing the correlation between the probability of severe/fatal injuries and impact velocity for ego vehicle, opponent bike and opponent passenger car. Side of opponent passenger car is separated into impact zones AC and B. The standard deviation is shown as an enclosing tube for each model.}
    \label{fig:regression_models}
\end{figure}

Separate models are created for the side crashes of each opponent type, i.e., bicycle and passenger car opponent, as well as for the front crash of the ego vehicle. For the latter, only crashes with other passenger car opponents are considered, as crashes with bicycles almost never result in severe or fatal injuries of the ego passenger car occupants. Furthermore, for the opponent passenger car, the side of the car is divided into three different impact zones, each stretching over one third of the opponent's side: Zone A, covering the front, zone B, containing the passenger cabin and zone C, covering the back of the vehicle. One model is created for zone B and a separate model for the combined zones A and C\footnote{Originally, separated models were created for zones A and C as well. However, GIDAS does not contain a sufficient amount of data for zone C to enable the regression process to converge to a meaningful result, which motivates the zone combination. The underlying assumption that crashes into zones A and C have a similar effect on the injury severity is confirmed by the low standard deviation of the resulting function fit.}. For the bicycle opponent, the final model does not contain zone differentiation, as first results showed no significant differences between the curves for respective zones. The final models  are shown in Figure~\ref{fig:regression_models}, together with the corresponding standard deviations.

The expected higher vulnerability of an opponent bicyclist can clearly be observed in the figure: The model exhibits all over higher values and a sharper rise in the curve compared to the car vehicle type. Already at rather low impact velocities of 20 kph, the probability of a severe/fatal crash lies above 20~\% and reaches almost 80~\% at an impact velocity of 60 kph. As it can be seen, the standard deviation of the model is almost zero, indicating a good model fit.


Comparing the different zones for the passenger car opponent, it is apparent that a crash into the middle third, i.e., the driver cabin, will more likely result in a severe/fatal injury compared to crashes in the front or rear third. For example, at 60 kph the probability lies at about 30~\% for crashes into the driver cabin, whereas crashes into the front or back will result in a severe/fatal accident with a probability below 20~\%. Compared to the bicycle model, the standard deviation is significantly higher with a value of 0.04 evaluated at 50 kph for zone B. This is partly due to the lower number of available crashes with 196 crashes for zone B and 423 crashes for combined zones A/C. Front crashes of the ego vehicle with other passenger cars are roughly comparable to crashes into the driver cabin of the opponent vehicle, exhibiting slightly lower numbers for low impact velocities while taking on higher values above 80 kph. Also, the standard deviation of the curve is reduced with a value of 0.017 at 50 kph. 


\section{Evaluation methodology}\label{sec:evaluationmethodology}
The benefit of using V2X for brake triggering in obstructed crossing situations is evaluated by comparing the performances of the V2X-enhanced 2-stage brake with the result for the AEB. Different use cases and scenarios are defined and simulated using the respective braking systems~\footnote{Although only a specific subset of crossing situations is evaluated in the subsequent analysis, the braking system can be directly applied to more complex, real world situations.}. The results are analysed using two key performance indicators (KPIs) that evaluate the \textit{crash avoidance} and \textit{crash mitigation} performances of the braking system under test.


\subsection{Use cases, scenarios and variations}\label{subsec:usecasesandscenarios}
\begin{figure}
    \centering
    \includegraphics[width=.6\linewidth]{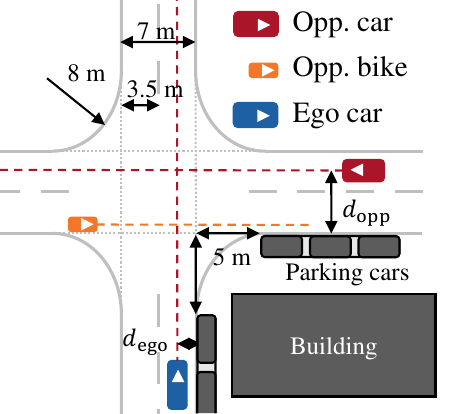}
    \caption{SCP use case crossing layout with ego and opponent vehicles, building and parking car view obstruction as well as distances $d_\mathrm{ego}$, $d_\mathrm{opp}$ from ego and opponent vehicle paths to view obstructions.}
    \label{fig:crossinglayout}
\end{figure}
 A \textit{use case} describes a general situation, in which the braking strategies under test are evaluated. Its definition includes the environment (e.g., an obstructed intersection), the involved vehicles (e.g., an ego and an opponent vehicle) and the vehicles paths (e.g., vehicles are crossing in a straight path from the left or right). 

 To ensure that the use cases analysed in this work are of importance in real-world crash situations, we rely for their definition on the outcome of the SECUR\footnote{SECUR (Safety Enhancement through Connected Users on the Road) was a European research project, in which different industrial stakeholders studied the potential of V2X to improve road user safety.} deliverable 3.1~\cite{SECUR_DEL31}. From these use cases, we adapt the obstructed crossing use cases SCP-RD Passenger Car, SCP-LD Passenger Car, SCP-RD Bicyclist and SCP-LD Bicyclist. SCP stands for straight crossing path and RD, LD for right direction, left direction, respectively, while passenger car and bicyclist describe the crossing opponent. 
 The basic layout of these use cases is depicted in Figure~\ref{fig:crossinglayout}: An ego vehicle and an opponent vehicle approach a crossing, consisting of intersecting roads with two lanes that cross each other with an angle of 90°. Both vehicles aim to cross the intersection in a straight line. The line of sight between the ego vehicle and the opponent vehicle is blocked by a view obstruction, which is assumed to be either parking cars or a building obstruction.
 
 For every use case, a variety of \textit{scenarios} are defined that differ from each other regarding changeable characteristics of the setting, including obstruction type and distances $d_\mathrm{ego}$, $d_\mathrm{opp}$ of vehicles paths to the obstruction. Two types of view obstructions, parking cars and building, are considered, where the former consist of two rows of cars parking at the road side, see Figure~\ref{fig:crossinglayout}. In total, 35 scenarios of the selected use cases are analysed. The parameters defining the scenarios are listed in Table~\ref{tab:param_scen}.
\begin{table}
    \centering
    \caption{Overview of scenarios associated with their respective use cases. *: distances $d_\mathrm{ego}$ and $d_\mathrm{opp}$ are varied in 0.5 m steps, combination yields nine different scenarios. **: Ego vehicle road is a one way street. ***Special case, where there is only a row of cars parallel to ego path and not opponent path.}
    \label{tab:param_scen}
    \begin{tabular}{l l | l l l}
       Use case   &  Scen. & Ob. type & $d_\mathrm{ego}$ in m & $d_\mathrm{opp}$ in m \\ \hline
        \multirow{2}{*}{SCP-RD PC}  & 1      & cars     & 1.925           & 5.425   \\
                  & 2 - 10*    & build.   & 3.25 - 4.25     & 6.75 - 7.75  \\\hline
      \multirow{2}{*}{SCP-LD PC} & 11 & cars & 5.425  & 1.925  \\
       & 12 - 20* & build. & 6.75 - 7.75  & 3.25  - 4.25  \\
       \multirow{2}{*}{one way**}& 21  & cars & 1.75  & 1.75  \\
       & 22 - 30* & build. & 3.25 - 4.25  & 3.25 - 4.25  \\ \hline
       \multirow{2}{*}{SCP-RD B} & 31 & build. & 4.2  & 2.7  \\ 
        & 32 & build. & 3.25  & 3.75  \\ \hline
       SCP-LD B & 33 & build.& 6.75  & 2.125   \\
       one way** & 34 & build. & 3.25  & 2.125  \\
       SCP-RD PC & 35 & cars*** & 1.65  & 20 
    \end{tabular}
\end{table}

\begin{figure}
    \centering
    \includegraphics[width=0.4\linewidth]{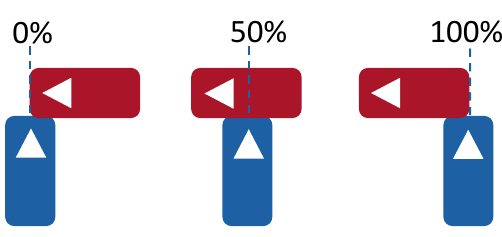}
    \caption{Depiction of unbraked impact position in percent of opponent length.}
    \label{fig:unbraked impact location}
\end{figure}

For every scenario of every use case, different \textit{variations} of the respective scenarios are defined. These variations include the initial velocities of the vehicles as well as the unbraked impact location. The latter describes the impact point, where the front of the ego vehicle hits the side of the opponent vehicle if no brake is applied. The concept is visualized in Figure~\ref{fig:unbraked impact location}. In Table~\ref{tab:variation}, the variation values are described. As every combination is tested, we receive 125 test cases for opponent cars and 75 test cases for opponent bicycles.

\begin{table}[]
    \centering
    \caption{Variation of initial velocity and impact location for every scenario.}
    \label{tab:variation}
    \begin{tabular}{r | l l }
                    &       init. velocity        &  impact loc.  \\ \hline
       opp. car    &  20, 30,..., 60 kph  & 0, 25, ..., 100~\% \\
      opp. bike &  5, 10, ..., 25 kph  & 0, 50, 100~\% \\
      ego car &  20, 30,..., 60 kph  & front
    \end{tabular}
    
\end{table}

\subsection{KPIs}\label{subsec:KPIs}
In order to quantify the performance of a braking system, key performance indicators (KPIs) are defined that translate the result of several simulation runs to specific numbers. 
In this work, two different KPIs are used:
\begin{itemize}
    \item number of avoided crashes in percent and
    \item probability for crash participants to experience a severe or fatal injury.
\end{itemize}
Evaluating the number of avoided crashes by simulation is straightforward. The probability for severe/fatal injuries is calculated by evaluating the impact velocity and impact location of the ego vehicle and using the respective models described in subsection~\ref{subsec:severitymodels}.
\\

\section{Simulation}\label{sec:simulation}
\subsection{Simulation environment}\label{subsec:simultionenvironments}
\begin{figure}[h]
    \centering
    \includegraphics[width=.45\linewidth]{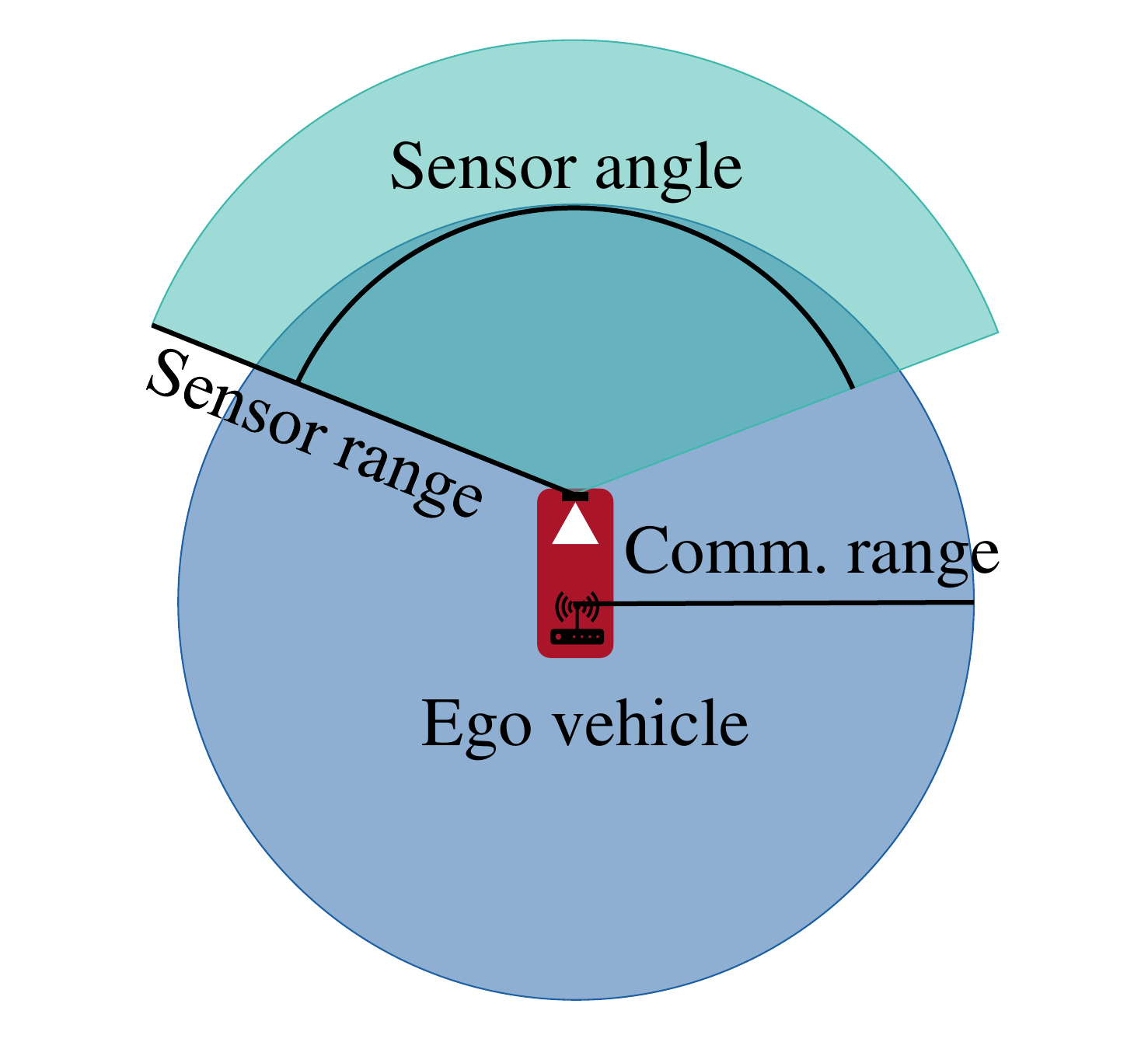}
    \caption{Ego vehicle with simplified sensor model, mounted at the vehicle front and defined by sensor angle and range, as well as communication system model, defined by communication range.}
    \label{fig:ego_vehicle_sensor_system}
\end{figure}

To simulate the vehicles movement in the use cases, described in the preceding section, a lightweight simulator is used. The dynamics of the vehicles are modelled according to the kinematic bicycle model. The vehicles themselves are represented as rectangles with a specific mass and center of gravity. Figure~\ref{fig:ego_vehicle_sensor_system} depicts how the onboard sensors and the communication system are modelled. As it can be observed, the onboard sensor system is simplified to a single sensor, which is attached to a mounting point at the front of the vehicle. The sensors field of view (FoV) is a circular segment, where the sensor range corresponds to the circle radius and the sensor angle to the spread of the segment. An opponent is detected whenever a specific point of the opponent's rectangle, the point of recognition, enters the sensor field of view. To account for the detection, classification and possible sensor fusion process, we assume a specific delay $\delta_\mathrm{class}$. 

The V2X communication is modelled as an additional sensor with a 360° sensor angle and a sensor range that is equivalent to a common inner-city communication range. The mount point of the V2X antenna is at the back of the vehicle. An opponent is recognized as soon as their antenna enters the field of view of the communication sensor. To account for the communication latency, decoding and processing a received message as well as for a possible sensor fusion, we assume a specific delay $\delta_\mathrm{comm}$. The delay $\delta_\mathrm{brake}$ takes the brake activation time into account. 

The simulation time is separated into 10 ms time slots. In each time slot, the sensor readings are updated and the brake decision process, depicted in Figure~\ref{fig:flow_diagram_twostage_brake}, is run. In the collision prediction process, trajectories of the ego vehicle and the opponent vehicle are calculated based on their current position and under constant velocity and heading assumptions. Based on the estimated trajectories, a crash is predicted if the vehicles occupy the same space. The result is passed to the braking decision process, where the triggering conditions, described in Subsection~\ref{subsec:twostagebrake}, are evaluated. 


The process is modelled detailed enough such that it is possible to make a valid assertion regarding the benefit of using V2X communication in the selected use cases, while factors such as false-positive brake activation and uncertainties regarding dynamic parameters are not considered. 

\subsection{Parameters}
In Table~\ref{tab:simulation_parameters}, all relevant brake parameters are summarized. Note that three different TTC thresholds, 1.25~s, 1.5~s and 2~s, are evaluated for the first stage brake. 
\begin{table}
    \centering
    \caption{Parameters for 1st stage and 2nd stage (AEB) brake. }
    \label{tab:simulation_parameters}
    \begin{tabular}{r|cc}
                               & 1st stage brake                  & 2nd stage brake/AEB  \\ \hline
         TTC threshold         & 1.25 s, 1.5 s, 2 s                        & 1.25 s \\
         max. decel.           & 4 $\mathrm{m/s^2}$              & 9 $\mathrm{m/s^2}$ \\
         Jerk                  & 45 $\mathrm{m/s^2}$              &  45 $\mathrm{m/s^2}$ \\
         delay detect.         & $\delta_\mathrm{comm} = 0.3 $ s & $\delta_\mathrm{class} = 0.2$ s \\
         delay application     & $\delta_\mathrm{brake}$ = 0.12 s & $\delta_\mathrm{brake}$ = 0.12 s
    \end{tabular}
\end{table}
Furthermore, three different onboard sensor sets are examined: a minimal sensor set, consisting of one camera sensor, a medium-class sensor set, consisting of one radar and one camera and a premium-class set, consisting of five radars and one camera. The sensor set models in the simulation differ from each other regarding the sensor angle and the mounting points, as shown in Table~\ref{tab:sensorsets}. 

\begin{table}
    \centering
     \caption{Overview of sensor sets used in simulation.}
    \label{tab:sensorsets}
    \begin{tabular}{r|cccc}
                 & Angle  & Range  & Mount point  & Detection\\ \hline
    V2X sensor   & 360°   & 56 m   &   3.75 m     & Antenna in FoV\\
    Minimal       & 100°   &  50 m  & 1.40 m       & 0.5 length in FoV                 \\
    Medium class  &  120°  &  50 m  & 0.25 m       & Front in FoV \\
    Premium      & 240°    &  50 m  & 0.25 m       & Front in FoV
    \end{tabular}
   
\end{table}

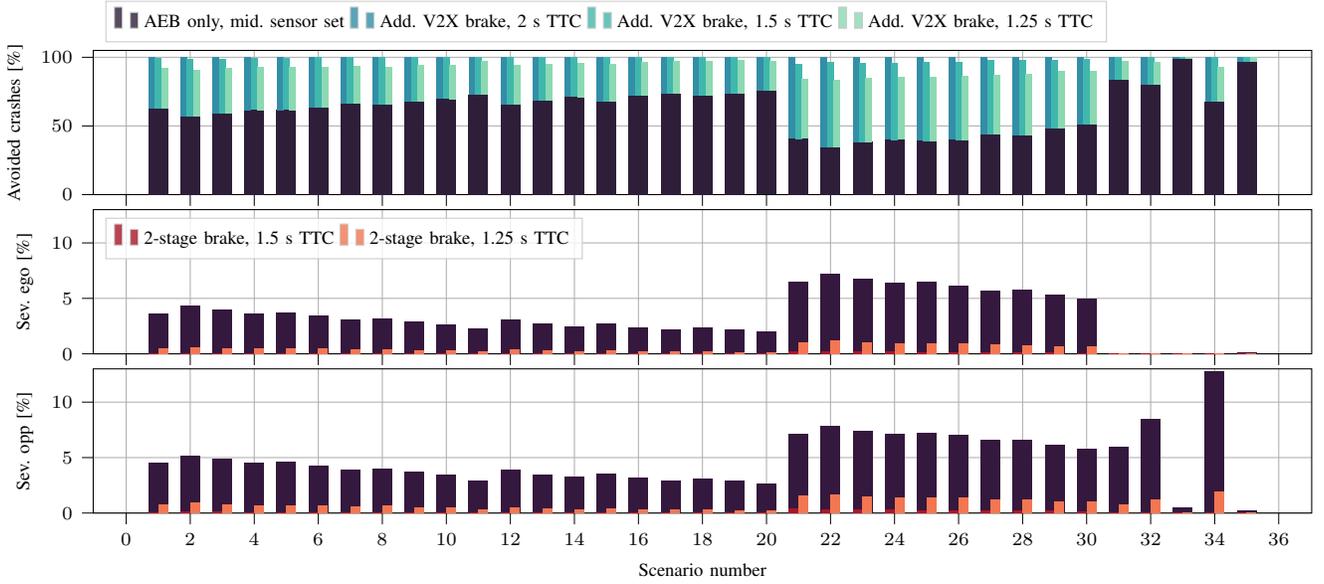
\begin{figure*}
    \centering
    \input{tikz/scen_bar_plot}
    \caption{Comparison between AEB-only and 2-stage brake (additional V2X brake) regarding remaining crashes (top plot), probability  for severe/fatal injuries for opponent (middle plot) and ego (bottom plot) over all 35 scenario. Scenarios 31 - 34 are bike opponent scenarios. Values are averaged over all variations of respective scenario and medium-class sensor set is used for AEB. Reduction means improvement }
    \label{fig:KPI_bar_plot}
\end{figure*}

\subsection{Simulation results}\label{subsec:simulationresults}
As described in Subsection~\ref{subsec:usecasesandscenarios}, we evaluate 35 different scenarios in total, each with 250 (150) variations of initial velocity and initial positions in opponent car (bike) scenarios. In total, 4175 cases were analysed, which are evaluated using the KPIs \textit{avoided crashes} and \textit{probability of severe/fatal injury}, earlier defined in Subsection~\ref{subsec:KPIs}.

\subsubsection{Overall results}
\begin{table}[h]
    \centering
    \caption{KPI evaluation of AEB-only and 2-stage brake simulations. Sev. ego and sev. opp describe the probability for the occupants of the respective vehicle to experience a severe/fatal injury.}
    \label{tab:results_AEB}
    \begin{tabular}{l|lll}
                    & Avoided crashes   & Sev. ego & Sev. opp      \\ \hline
       min - AEB    & 38.97~\%         & 5.74~\%  & 7.84~\%       \\
       mid - AEB    & 61.75~\%         & 3.45~\%  &  4.90~\%      \\
       prem - AEB   & 64.48~\%         & 3.34~\%  & 4.70~\%       \\ \hline
       min - 2 s    & 100~\%           & 0~\%     & 0~\%          \\
       mid - 2 s    & 100~\%           & 0~\%     &  0~\%         \\
       prem - 2 s   & 100~\%           & 0~\%     & 0~\%          \\ \hline
       min - 1.5 s  & 96.83~\%         & 0.16~\%  & 0.30~\%       \\
       mid - 1.5 s  & 98.87~\%          & 0.05~\%  &  0.09~\%      \\
       prem - 1.5 s & 98.87~\%         & 0.05~\%  & 0.09~\%       \\ \hline
       min - 1.25 s & 87.47~\%         & 0.80~\%  & 1.38~\%       \\
       mid - 1.25 s & 92.41~\%          & 0.46~\%  &  0.78~\%      \\
       prem - 1.25 s& 92.41~\%         & 0.46~\%  & 0.78~\%       \\ 
    \end{tabular}
    
\end{table}
In Table~\ref{tab:results_AEB}, the simulation results regarding all 35 scenarios with respective variations are listed. The KPI values, listed in the columns, are averaged over all 35 scenarios and respective variations. The first three rows depict the AEB-only results for the three sensor sets under test (see Table~\ref{tab:sensorsets}). The subsequent rows show the results for V2X-enhanced 2-stage brake for the TTC thresholds 2 s, 1.5 s and 1.25 s. 

Considering the AEB-only results, it can be observed that there is a significant performance difference between the minimal and the medium-class sensor set: Using the medium-class set, roughly 23~\% more crashes could be avoided compared to the minimal set. Furthermore, the probabilities for severe/fatal injuries are reduced by roughly 2.3~\% for the ego vehicle and 2.9~\% for the opponent vehicle. This performance improvement is due to two reasons. First, a sensor mount point closer to the front of the vehicle (see Table~\ref{tab:sensorsets}) enables an earlier recognition of the opponent before the crash. Second, the medium-class set has a wider field of view due to a greater sensor angle. The differences between the medium and premium-class sensor sets are far less distinctive: only about 2.7~\% more crashes can be avoided and the crash severity reduction is insignificant. As the sensor mount points are the same for both sets, the wider sensor angle of the premium set 
provides no advantage in the analysed use cases.

Using the 2-stage brake with a 2 s TTC threshold, all crashes can be avoided. However, reducing the TTC threshold to 1.25 s, 12.5~\% and 7.6~\% of crashes cannot be avoided for the minimal and the premium-class sensor sets, respectively. At the same time, the probability for severe/fatal injuries rises to a highest value of 1.38~\%, still rather low compared to the 7.84~\% of the AEB-only result. Reducing the TTC threshold from 2 to 1.5 s results in an insignificant performance degradation, especially for the medium and premium-class sensor set. Note that for the 2-stage brake, no performance differences can be observed between the medium and premium-class sensor sets.

Three main insights can be gathered from discussion of the overall results. For one, the highest improvement regarding the AEB performance can be achieved by using the medium sensor set compared to the minimal set, i.e., adding a radar to the video sensor. Second, it is possible to reduce the number of crashes to zero by adding the V2X-triggered brake in combination with all analysed onboard sensor sets, provided that the brake can be triggered 2 s before the crash. And third, reducing the TTC threshold of the V2X-triggered brake to 1.5 s increases the number of crashes and respective severity probabilities only insignificantly for the medium and premium-class sensor sets. 


\subsubsection{Scenario specific results}
Figure~\ref{fig:KPI_bar_plot} depicts the KPIs split over all 35 scenarios, averaged over the respective velocity and impact location variation of the respective scenario. All results are produced under usage of the medium-class onboard sensor set. The uppermost plot shows the crashes avoided by the AEB brake as the dark, broad bars, while the crashes that can be avoided only by use of the additional V2X brake are depicted for the TTC thresholds 2 s, 1.5 s and 1.25 s as slender bars in different hues of green. The combination of the dark, broad bar and the respective slender, green bar is the performance of the 2-stage brake. The middle and lower plot depict the probability of severe/fatal injury of the ego and opponent vehicles, respectively. Here, the results of the AEB-only brake, again illustrated as dark, broad bars, are compared against the 2-stage brake performance with two different TTC thresholds\footnote{As with a 2 s TTC threshold all crashes can be avoided, the severity probability is zero for all scenarios and is therefore not depicted.}, 1.5 s and 1.25 s, colored in hues of red. Note that the probabilities for the 1.5 s threshold are often zero or very low. Furthermore, scenarios 31 - 34 are bicycle opponent use cases, see Table~\ref{tab:param_scen}, wherefore the probabilities for severe/fatal crashes for the ego vehicle are zero for these scenarios.

It can be observed that the performance differences between the 2-stage brake at 2.0 s and 1.5 s TTC are almost negligible for the scenarios 1 - 5 (highest difference 1.6~\%) and non existent for scenarios 6 - 20 and 31 - 35. Only for the scenarios 21 - 30 significant differences can be observed. These scenarios also mark the worst performances for AEB and 2-stage brake regarding crash avoidance performance as well as ego severity probability, except the 2-stage brake with 2 s TTC, for which all crashes can be avoided. As it can be obtained from Table~\ref{tab:param_scen}, all these scenarios are one way left crossing use cases with a passenger car opponent. Due to this setup, the distances $d_\mathrm{ego}$ and $d_\mathrm{opp}$, defined in Figure~\ref{fig:crossinglayout}, are the smallest for all tested use cases, except for some bicycle use scenarios. Resulting from these close distances, the opponent vehicle can only be recognized shortly before the crash, which does not allow the AEB to prevent the crash. This late reaction cannot be fully compensated for by the V2X-triggered first stage brake of the 2-stage braking systems due to the limited braking force of 4 $\mathrm{m/s^2}$ for TTC thresholds of 1.25 s and 1.5 s. However, allowing the first stage brake to be triggered at 2 s TTC, the crashes for all variations are completely avoided. While the probability for a severe/fatal injury for the scenarios 21 - 30 is quite low for the 2-stage brake under 1.5 s TTC threshold (highest value 0.4~\% for opponent for scenario 21), the probability rises to a maximum of 1.69~\% for the 1.25 s TTC threshold.  

The scenarios 31 - 34 are bicycle opponent use cases. The crash avoidance performance of the AEB is comparatively good. Especially in scenario 33 almost all crashes are avoided by the AEB alone, due to a higher distance of the ego vehicle to the view obstruction (see Table~\ref{tab:param_scen}) in combination with the smaller target length of the bicycle compared to a passenger car. However, if a crash occurs, the probability of a severe/fatal injury of the bicycle opponent is high and peaks at 12.75~\% for scenario 34. This is due to the increased vulnerability of the bicycle riders, as shown by the respective severity model in Figure~\ref{fig:regression_models}.

In summary, it can be stressed again that by using an additional V2X-triggered brake, the crash avoidance and mitigation performance of an automatic braking system is increased dramatically in obstructed view use cases, especially when the distances to the obstruction are low, ensuing that the AEB cannot be triggered in time. The reduction of the severity probability is highest for the bicycle opponent scenarios, where already a V2X brake with a TTC trigger time of 1.5 s can avoid all crashes.


\section{Conclusion}\label{sec:conclusion}
The simulation results clearly display the high benefit of using V2X in an automated braking system in crossing situations, where an obstruction prevents the timely detection of the opponent vehicle by driver and onboard sensor systems. By using a 2 s TTC threshold for the V2X brake, crashes can be avoided completely in all tested cases. Even when reducing the TTC to 1.5 s, roughly 99~\% of crashes can be avoided, while simultaneously the risk of a false positive brake activation is reduced compared to the 2 s TTC trigger. An in-depth study regarding occurrence and consequences of false-positive brake activation is kept for future work.

Furthermore, real world implementations of the 2-stage brake will need to consider non-ideal assumptions regarding sensor equipment and V2X communication, such as uncertainties in dynamical parameters, i.e., position and velocity, as well as stochastic communication channels.


\addtolength{\textheight}{-12cm}   






%
%
%
\bibliographystyle{IEEEtran} 
\input{root.bbl}


\end{document}

%% file: tikz/twostagebrake_flow.tex
\begin{tikzpicture}

\definecolor{firebrick1702141}{RGB}{214,39,40}
\definecolor{darkorange}{RGB}{255,127,14}
\definecolor{steelblue}{RGB}{30,96,164}
\definecolor{lightseagreen64183173}{HTML}{2CA02C}
\definecolor{purple}{HTML}{9467BD}
\definecolor{firebrick1702141}{RGB}{170,21,41}
\definecolor{lightseagreen6418317364183173}{RGB}{64,183,173}

\tikzstyle{my below of} = [below=of #1.south]
\tikzstyle{dist below of} = [below=of #1.south, node distance = 1cm]
\tikzstyle{my right of} = [right=of #1.east]
\tikzstyle{my left of} = [left=of #1.west]
\tikzstyle{my above of} = [above=of #1.north]

\tikzstyle{arrow} = [thick,->,>=stealth]

\tikzstyle{interface} = [rectangle, rounded corners, minimum width=1cm, minimum height=0.4cm,text centered, draw=black, fill=firebrick1702141!30, font=\scriptsize]
\tikzstyle{function} = [rectangle, minimum width=3.5cm, minimum height=0.75cm,text centered, draw=black, fill=lightseagreen64183173!50, font=\scriptsize]
\tikzstyle{generic} = [rectangle, minimum width=3.5cm, minimum height=0.75cm,text centered, draw=black, fill=purple!30, font=\scriptsize]
\tikzstyle{AEB} = [rectangle, minimum width=3.5cm, minimum height=4.5cm,text centered, draw=black, fill=steelblue!30, font=\scriptsize]
\tikzstyle{textblock} = [rectangle,text centered, font=\scriptsize]

\node (v2xobjects) [interface] {V2X objects};
\node (egostate) [interface, right of=v2xobjects, node distance=2.5cm] {Ego state};
\node (sensorobjects) [interface, right of=egostate, node distance=2.5cm] {Sensor objects};


\node (1ststage) [textblock,  dist below of=v2xobjects] {1st stage \    \ \ \ \ \ \ \ \ \ \    partial brake};
\node (fusion) [function,  below of=1ststage, align=center, node distance = 0.75cm] {Fusion/\\ scene understanding};
\node (collisionpred) [function,  below of=fusion, node distance = 1cm] {Collision prediction};
\node (brakingdecision) [function,  below of=collisionpred, node distance = 1cm] {Braking decision};
\node (driverveto) [function,  below of=brakingdecision, node distance = 1cm] {Driver override};
\scoped[on background layer]{\node [fit=(1ststage)(driverveto), fill=lightseagreen64183173!30, draw=black] {2nd stage};}

\node (aeb) [AEB, dist below of=sensorobjects, align=center, node distance = 2cm] {2nd stage: AEB};

\node (interventionman) [generic, below  of=egostate, node distance = 7cm] {Intervention manager};

\draw [arrow] node{} (-0.25cm,-0.22)  -- node{} (-0.25cm,-1.85cm);

\draw [arrow]  (egostate.south) -- node{} (2.5,-0.5cm) -- node{} (4.75cm,-0.5cm) -- node{} (4.75cm,-1.25cm) ;
\draw [arrow] (egostate.south) -- node{} (2.5cm,-0.5cm) -- node{} (0cm,-0.5cm) -- node{} (0cm,-1.85cm);

\draw [arrow] (sensorobjects.south) -- (aeb.north);
\draw [arrow] (sensorobjects.south) -- node{} (5,-0.75cm) -- node{} (0.25,-0.75cm) -- node{} (0.25cm,-1.85cm);

\draw [arrow] (fusion.south) -- (collisionpred.north);
\draw [arrow] (collisionpred) -- (brakingdecision);
\draw [arrow] (brakingdecision) -- (driverveto);

\draw [arrow] (aeb.south) -- node{} (5,-6cm) -- node{} (3,-6cm) -- node{} (3,-6.64cm);
\draw [arrow] (driverveto.south) -- node{} (0,-6cm) -- node{} (2,-6cm) -- node{} (2,-6.64cm);

\end{tikzpicture}

%% file: tikz/scen_bar_plot.tex
\begin{tikzpicture}

\definecolor{burlywood246180142}{RGB}{246,180,142}
\definecolor{coral24311881}{RGB}{243,118,81}
\definecolor{darkgray176}{RGB}{176,176,176}
\definecolor{darkslategray463058}{RGB}{46,30,58}
\definecolor{darkslategray532461}{RGB}{53,24,61}
\definecolor{lightgray204}{RGB}{204,204,204}
\definecolor{lightseagreen64183173}{RGB}{64,183,173}
\definecolor{mediumaquamarine139217178}{RGB}{139,217,178}
\definecolor{steelblue51142167}{RGB}{51,142,167}
\definecolor{firebrick1702141}{RGB}{170,21,41}

\begin{groupplot}[group style={group size=1 by 3, vertical sep=0.2cm}]
\nextgroupplot[
legend cell align={left},
legend style={
  fill opacity=0.8,
  legend columns=4,
  draw opacity=1,
  text opacity=1,
  at={(0.01,1.2)},
  anchor=west,
  draw=lightgray204
},
tick align=outside,
tick pos=left,
xticklabels={},
x grid style={darkgray176},
xmajorgrids,
xmin=-1.03, xmax=37.03,
xtick style={color=black},
y grid style={darkgray176},
ylabel={Avoided crashes [\%]},
ymajorgrids,
ymin=0, ymax=105,
ytick style={color=black},
height = 3.5cm,
width = \textwidth,
font=\scriptsize
]
\draw[draw=none,fill=darkslategray463058] (axis cs:0.7,0) rectangle (axis cs:1.3,62.4);
\addlegendimage{ybar,ybar legend,draw=none,fill=darkslategray463058}
\addlegendentry{AEB only, mid. sensor set}

\draw[draw=none,fill=darkslategray463058] (axis cs:1.7,0) rectangle (axis cs:2.3,56.8);
\draw[draw=none,fill=darkslategray463058] (axis cs:2.7,0) rectangle (axis cs:3.3,59.2);
\draw[draw=none,fill=darkslategray463058] (axis cs:3.7,0) rectangle (axis cs:4.3,61.6);
\draw[draw=none,fill=darkslategray463058] (axis cs:4.7,0) rectangle (axis cs:5.3,61.6);
\draw[draw=none,fill=darkslategray463058] (axis cs:5.7,0) rectangle (axis cs:6.3,63.2);
\draw[draw=none,fill=darkslategray463058] (axis cs:6.7,0) rectangle (axis cs:7.3,66.4);
\draw[draw=none,fill=darkslategray463058] (axis cs:7.7,0) rectangle (axis cs:8.3,65.6);
\draw[draw=none,fill=darkslategray463058] (axis cs:8.7,0) rectangle (axis cs:9.3,68);
\draw[draw=none,fill=darkslategray463058] (axis cs:9.7,0) rectangle (axis cs:10.3,69.6);
\draw[draw=none,fill=darkslategray463058] (axis cs:10.7,0) rectangle (axis cs:11.3,72.8);
\draw[draw=none,fill=darkslategray463058] (axis cs:11.7,0) rectangle (axis cs:12.3,65.6);
\draw[draw=none,fill=darkslategray463058] (axis cs:12.7,0) rectangle (axis cs:13.3,68.8);
\draw[draw=none,fill=darkslategray463058] (axis cs:13.7,0) rectangle (axis cs:14.3,71.2);
\draw[draw=none,fill=darkslategray463058] (axis cs:14.7,0) rectangle (axis cs:15.3,68);
\draw[draw=none,fill=darkslategray463058] (axis cs:15.7,0) rectangle (axis cs:16.3,72);
\draw[draw=none,fill=darkslategray463058] (axis cs:16.7,0) rectangle (axis cs:17.3,73.6);
\draw[draw=none,fill=darkslategray463058] (axis cs:17.7,0) rectangle (axis cs:18.3,72);
\draw[draw=none,fill=darkslategray463058] (axis cs:18.7,0) rectangle (axis cs:19.3,73.6);
\draw[draw=none,fill=darkslategray463058] (axis cs:19.7,0) rectangle (axis cs:20.3,76);
\draw[draw=none,fill=darkslategray463058] (axis cs:20.7,0) rectangle (axis cs:21.3,40.8);
\draw[draw=none,fill=darkslategray463058] (axis cs:21.7,0) rectangle (axis cs:22.3,34.4);
\draw[draw=none,fill=darkslategray463058] (axis cs:22.7,0) rectangle (axis cs:23.3,38.4);
\draw[draw=none,fill=darkslategray463058] (axis cs:23.7,0) rectangle (axis cs:24.3,40);
\draw[draw=none,fill=darkslategray463058] (axis cs:24.7,0) rectangle (axis cs:25.3,39.2);
\draw[draw=none,fill=darkslategray463058] (axis cs:25.7,0) rectangle (axis cs:26.3,40);
\draw[draw=none,fill=darkslategray463058] (axis cs:26.7,0) rectangle (axis cs:27.3,44);
\draw[draw=none,fill=darkslategray463058] (axis cs:27.7,0) rectangle (axis cs:28.3,43.2);
\draw[draw=none,fill=darkslategray463058] (axis cs:28.7,0) rectangle (axis cs:29.3,48);
\draw[draw=none,fill=darkslategray463058] (axis cs:29.7,0) rectangle (axis cs:30.3,51.2);
\draw[draw=none,fill=darkslategray463058] (axis cs:30.7,0) rectangle (axis cs:31.3,84);
\draw[draw=none,fill=darkslategray463058] (axis cs:31.7,0) rectangle (axis cs:32.3,80);
\draw[draw=none,fill=darkslategray463058] (axis cs:32.7,0) rectangle (axis cs:33.3,98.6666666666667);
\draw[draw=none,fill=darkslategray463058] (axis cs:33.7,0) rectangle (axis cs:34.3,68);
\draw[draw=none,fill=darkslategray463058] (axis cs:34.7,0) rectangle (axis cs:35.3,96.8);
\draw[draw=none,fill=steelblue51142167] (axis cs:0.7,62.4) rectangle (axis cs:0.9,100);
\addlegendimage{ybar,ybar legend,draw=none,fill=steelblue51142167}
\addlegendentry{Add. V2X brake, 2 s TTC}

\draw[draw=none,fill=steelblue51142167] (axis cs:1.7,56.8) rectangle (axis cs:1.9,100);
\draw[draw=none,fill=steelblue51142167] (axis cs:2.7,59.2) rectangle (axis cs:2.9,100);
\draw[draw=none,fill=steelblue51142167] (axis cs:3.7,61.6) rectangle (axis cs:3.9,100);
\draw[draw=none,fill=steelblue51142167] (axis cs:4.7,61.6) rectangle (axis cs:4.9,100);
\draw[draw=none,fill=steelblue51142167] (axis cs:5.7,63.2) rectangle (axis cs:5.9,100);
\draw[draw=none,fill=steelblue51142167] (axis cs:6.7,66.4) rectangle (axis cs:6.9,100);
\draw[draw=none,fill=steelblue51142167] (axis cs:7.7,65.6) rectangle (axis cs:7.9,100);
\draw[draw=none,fill=steelblue51142167] (axis cs:8.7,68) rectangle (axis cs:8.9,100);
\draw[draw=none,fill=steelblue51142167] (axis cs:9.7,69.6) rectangle (axis cs:9.9,100);
\draw[draw=none,fill=steelblue51142167] (axis cs:10.7,72.8) rectangle (axis cs:10.9,100);
\draw[draw=none,fill=steelblue51142167] (axis cs:11.7,65.6) rectangle (axis cs:11.9,100);
\draw[draw=none,fill=steelblue51142167] (axis cs:12.7,68.8) rectangle (axis cs:12.9,100);
\draw[draw=none,fill=steelblue51142167] (axis cs:13.7,71.2) rectangle (axis cs:13.9,100);
\draw[draw=none,fill=steelblue51142167] (axis cs:14.7,68) rectangle (axis cs:14.9,100);
\draw[draw=none,fill=steelblue51142167] (axis cs:15.7,72) rectangle (axis cs:15.9,100);
\draw[draw=none,fill=steelblue51142167] (axis cs:16.7,73.6) rectangle (axis cs:16.9,100);
\draw[draw=none,fill=steelblue51142167] (axis cs:17.7,72) rectangle (axis cs:17.9,100);
\draw[draw=none,fill=steelblue51142167] (axis cs:18.7,73.6) rectangle (axis cs:18.9,100);
\draw[draw=none,fill=steelblue51142167] (axis cs:19.7,76) rectangle (axis cs:19.9,100);
\draw[draw=none,fill=steelblue51142167] (axis cs:20.7,40.8) rectangle (axis cs:20.9,100);
\draw[draw=none,fill=steelblue51142167] (axis cs:21.7,34.4) rectangle (axis cs:21.9,100);
\draw[draw=none,fill=steelblue51142167] (axis cs:22.7,38.4) rectangle (axis cs:22.9,100);
\draw[draw=none,fill=steelblue51142167] (axis cs:23.7,40) rectangle (axis cs:23.9,100);
\draw[draw=none,fill=steelblue51142167] (axis cs:24.7,39.2) rectangle (axis cs:24.9,100);
\draw[draw=none,fill=steelblue51142167] (axis cs:25.7,40) rectangle (axis cs:25.9,100);
\draw[draw=none,fill=steelblue51142167] (axis cs:26.7,44) rectangle (axis cs:26.9,100);
\draw[draw=none,fill=steelblue51142167] (axis cs:27.7,43.2) rectangle (axis cs:27.9,100);
\draw[draw=none,fill=steelblue51142167] (axis cs:28.7,48) rectangle (axis cs:28.9,100);
\draw[draw=none,fill=steelblue51142167] (axis cs:29.7,51.2) rectangle (axis cs:29.9,100);
\draw[draw=none,fill=steelblue51142167] (axis cs:30.7,84) rectangle (axis cs:30.9,100);
\draw[draw=none,fill=steelblue51142167] (axis cs:31.7,80) rectangle (axis cs:31.9,100);
\draw[draw=none,fill=steelblue51142167] (axis cs:32.7,98.6666666666667) rectangle (axis cs:32.9,100);
\draw[draw=none,fill=steelblue51142167] (axis cs:33.7,68) rectangle (axis cs:33.9,100);
\draw[draw=none,fill=steelblue51142167] (axis cs:34.7,96.8) rectangle (axis cs:34.9,100);
\draw[draw=none,fill=lightseagreen64183173] (axis cs:0.9,62.4) rectangle (axis cs:1.1,99.2);
\addlegendimage{ybar,ybar legend,draw=none,fill=lightseagreen64183173}
\addlegendentry{Add. V2X brake, 1.5 s TTC}

\draw[draw=none,fill=lightseagreen64183173] (axis cs:1.9,56.8) rectangle (axis cs:2.1,98.4);
\draw[draw=none,fill=lightseagreen64183173] (axis cs:2.9,59.2) rectangle (axis cs:3.1,98.4);
\draw[draw=none,fill=lightseagreen64183173] (axis cs:3.9,61.6) rectangle (axis cs:4.1,99.2);
\draw[draw=none,fill=lightseagreen64183173] (axis cs:4.9,61.6) rectangle (axis cs:5.1,99.2);
\draw[draw=none,fill=lightseagreen64183173] (axis cs:5.9,63.2) rectangle (axis cs:6.1,100);
\draw[draw=none,fill=lightseagreen64183173] (axis cs:6.9,66.4) rectangle (axis cs:7.1,100);
\draw[draw=none,fill=lightseagreen64183173] (axis cs:7.9,65.6) rectangle (axis cs:8.1,100);
\draw[draw=none,fill=lightseagreen64183173] (axis cs:8.9,68) rectangle (axis cs:9.1,100);
\draw[draw=none,fill=lightseagreen64183173] (axis cs:9.9,69.6) rectangle (axis cs:10.1,100);
\draw[draw=none,fill=lightseagreen64183173] (axis cs:10.9,72.8) rectangle (axis cs:11.1,100);
\draw[draw=none,fill=lightseagreen64183173] (axis cs:11.9,65.6) rectangle (axis cs:12.1,100);
\draw[draw=none,fill=lightseagreen64183173] (axis cs:12.9,68.8) rectangle (axis cs:13.1,100);
\draw[draw=none,fill=lightseagreen64183173] (axis cs:13.9,71.2) rectangle (axis cs:14.1,100);
\draw[draw=none,fill=lightseagreen64183173] (axis cs:14.9,68) rectangle (axis cs:15.1,100);
\draw[draw=none,fill=lightseagreen64183173] (axis cs:15.9,72) rectangle (axis cs:16.1,100);
\draw[draw=none,fill=lightseagreen64183173] (axis cs:16.9,73.6) rectangle (axis cs:17.1,100);
\draw[draw=none,fill=lightseagreen64183173] (axis cs:17.9,72) rectangle (axis cs:18.1,100);
\draw[draw=none,fill=lightseagreen64183173] (axis cs:18.9,73.6) rectangle (axis cs:19.1,100);
\draw[draw=none,fill=lightseagreen64183173] (axis cs:19.9,76) rectangle (axis cs:20.1,100);
\draw[draw=none,fill=lightseagreen64183173] (axis cs:20.9,40.8) rectangle (axis cs:21.1,95.2);
\draw[draw=none,fill=lightseagreen64183173] (axis cs:21.9,34.4) rectangle (axis cs:22.1,96);
\draw[draw=none,fill=lightseagreen64183173] (axis cs:22.9,38.4) rectangle (axis cs:23.1,96);
\draw[draw=none,fill=lightseagreen64183173] (axis cs:23.9,40) rectangle (axis cs:24.1,96);
\draw[draw=none,fill=lightseagreen64183173] (axis cs:24.9,39.2) rectangle (axis cs:25.1,96.8);
\draw[draw=none,fill=lightseagreen64183173] (axis cs:25.9,40) rectangle (axis cs:26.1,96.8);
\draw[draw=none,fill=lightseagreen64183173] (axis cs:26.9,44) rectangle (axis cs:27.1,97.6);
\draw[draw=none,fill=lightseagreen64183173] (axis cs:27.9,43.2) rectangle (axis cs:28.1,97.6);
\draw[draw=none,fill=lightseagreen64183173] (axis cs:28.9,48) rectangle (axis cs:29.1,97.6);
\draw[draw=none,fill=lightseagreen64183173] (axis cs:29.9,51.2) rectangle (axis cs:30.1,98.4);
\draw[draw=none,fill=lightseagreen64183173] (axis cs:30.9,84) rectangle (axis cs:31.1,100);
\draw[draw=none,fill=lightseagreen64183173] (axis cs:31.9,80) rectangle (axis cs:32.1,100);
\draw[draw=none,fill=lightseagreen64183173] (axis cs:32.9,98.6666666666667) rectangle (axis cs:33.1,100);
\draw[draw=none,fill=lightseagreen64183173] (axis cs:33.9,68) rectangle (axis cs:34.1,100);
\draw[draw=none,fill=lightseagreen64183173] (axis cs:34.9,96.8) rectangle (axis cs:35.1,100);
\draw[draw=none,fill=mediumaquamarine139217178] (axis cs:1.1,62.4) rectangle (axis cs:1.3,92);
\addlegendimage{ybar,ybar legend,draw=none,fill=mediumaquamarine139217178}
\addlegendentry{Add. V2X brake, 1.25 s TTC}

\draw[draw=none,fill=mediumaquamarine139217178] (axis cs:2.1,56.8) rectangle (axis cs:2.3,90.4);
\draw[draw=none,fill=mediumaquamarine139217178] (axis cs:3.1,59.2) rectangle (axis cs:3.3,92);
\draw[draw=none,fill=mediumaquamarine139217178] (axis cs:4.1,61.6) rectangle (axis cs:4.3,92.8);
\draw[draw=none,fill=mediumaquamarine139217178] (axis cs:5.1,61.6) rectangle (axis cs:5.3,92.8);
\draw[draw=none,fill=mediumaquamarine139217178] (axis cs:6.1,63.2) rectangle (axis cs:6.3,92.8);
\draw[draw=none,fill=mediumaquamarine139217178] (axis cs:7.1,66.4) rectangle (axis cs:7.3,93.6);
\draw[draw=none,fill=mediumaquamarine139217178] (axis cs:8.1,65.6) rectangle (axis cs:8.3,92.8);
\draw[draw=none,fill=mediumaquamarine139217178] (axis cs:9.1,68) rectangle (axis cs:9.3,94.4);
\draw[draw=none,fill=mediumaquamarine139217178] (axis cs:10.1,69.6) rectangle (axis cs:10.3,94.4);
\draw[draw=none,fill=mediumaquamarine139217178] (axis cs:11.1,72.8) rectangle (axis cs:11.3,96.8);
\draw[draw=none,fill=mediumaquamarine139217178] (axis cs:12.1,65.6) rectangle (axis cs:12.3,94.4);
\draw[draw=none,fill=mediumaquamarine139217178] (axis cs:13.1,68.8) rectangle (axis cs:13.3,95.2);
\draw[draw=none,fill=mediumaquamarine139217178] (axis cs:14.1,71.2) rectangle (axis cs:14.3,96);
\draw[draw=none,fill=mediumaquamarine139217178] (axis cs:15.1,68) rectangle (axis cs:15.3,95.2);
\draw[draw=none,fill=mediumaquamarine139217178] (axis cs:16.1,72) rectangle (axis cs:16.3,96);
\draw[draw=none,fill=mediumaquamarine139217178] (axis cs:17.1,73.6) rectangle (axis cs:17.3,96.8);
\draw[draw=none,fill=mediumaquamarine139217178] (axis cs:18.1,72) rectangle (axis cs:18.3,96.8);
\draw[draw=none,fill=mediumaquamarine139217178] (axis cs:19.1,73.6) rectangle (axis cs:19.3,97.6);
\draw[draw=none,fill=mediumaquamarine139217178] (axis cs:20.1,76) rectangle (axis cs:20.3,97.6);
\draw[draw=none,fill=mediumaquamarine139217178] (axis cs:21.1,40.8) rectangle (axis cs:21.3,84);
\draw[draw=none,fill=mediumaquamarine139217178] (axis cs:22.1,34.4) rectangle (axis cs:22.3,83.2);
\draw[draw=none,fill=mediumaquamarine139217178] (axis cs:23.1,38.4) rectangle (axis cs:23.3,84.8);
\draw[draw=none,fill=mediumaquamarine139217178] (axis cs:24.1,40) rectangle (axis cs:24.3,85.6);
\draw[draw=none,fill=mediumaquamarine139217178] (axis cs:25.1,39.2) rectangle (axis cs:25.3,85.6);
\draw[draw=none,fill=mediumaquamarine139217178] (axis cs:26.1,40) rectangle (axis cs:26.3,86.4);
\draw[draw=none,fill=mediumaquamarine139217178] (axis cs:27.1,44) rectangle (axis cs:27.3,87.2);
\draw[draw=none,fill=mediumaquamarine139217178] (axis cs:28.1,43.2) rectangle (axis cs:28.3,88);
\draw[draw=none,fill=mediumaquamarine139217178] (axis cs:29.1,48) rectangle (axis cs:29.3,89.6);
\draw[draw=none,fill=mediumaquamarine139217178] (axis cs:30.1,51.2) rectangle (axis cs:30.3,89.6);
\draw[draw=none,fill=mediumaquamarine139217178] (axis cs:31.1,84) rectangle (axis cs:31.3,97.3333333333333);
\draw[draw=none,fill=mediumaquamarine139217178] (axis cs:32.1,80) rectangle (axis cs:32.3,96);
\draw[draw=none,fill=mediumaquamarine139217178] (axis cs:33.1,98.6666666666667) rectangle (axis cs:33.3,100);
\draw[draw=none,fill=mediumaquamarine139217178] (axis cs:34.1,68) rectangle (axis cs:34.3,93.3333333333333);
\draw[draw=none,fill=mediumaquamarine139217178] (axis cs:35.1,96.8) rectangle (axis cs:35.3,100);

\nextgroupplot[
legend cell align={left},
legend style={
  fill opacity=0.8,
  legend columns=3,
  draw opacity=1,
  text opacity=1,
  at={(0.01,0.8)},
  anchor= west,
  draw=lightgray204
},
tick align=outside,
tick pos=left,
xticklabels={},
x grid style={darkgray176},
xmajorgrids,
xmin=-1.03, xmax=37.03,
xtick style={color=black},
y grid style={darkgray176},
ylabel={Sev. ego [\%]},
ymajorgrids,
ymin=0, ymax=13,
ytick style={color=black},
height = 3.5cm,
width = \textwidth,
font=\scriptsize
]
\draw[draw=none,fill=darkslategray532461] (axis cs:0.7,0) rectangle (axis cs:1.3,3.67);
\draw[draw=none,fill=darkslategray532461] (axis cs:1.7,0) rectangle (axis cs:2.3,4.33);
\draw[draw=none,fill=darkslategray532461] (axis cs:2.7,0) rectangle (axis cs:3.3,4.02);
\draw[draw=none,fill=darkslategray532461] (axis cs:3.7,0) rectangle (axis cs:4.3,3.65);
\draw[draw=none,fill=darkslategray532461] (axis cs:4.7,0) rectangle (axis cs:5.3,3.71);
\draw[draw=none,fill=darkslategray532461] (axis cs:5.7,0) rectangle (axis cs:6.3,3.44);
\draw[draw=none,fill=darkslategray532461] (axis cs:6.7,0) rectangle (axis cs:7.3,3.11);
\draw[draw=none,fill=darkslategray532461] (axis cs:7.7,0) rectangle (axis cs:8.3,3.22);
\draw[draw=none,fill=darkslategray532461] (axis cs:8.7,0) rectangle (axis cs:9.3,2.88);
\draw[draw=none,fill=darkslategray532461] (axis cs:9.7,0) rectangle (axis cs:10.3,2.66);
\draw[draw=none,fill=darkslategray532461] (axis cs:10.7,0) rectangle (axis cs:11.3,2.3);
\draw[draw=none,fill=darkslategray532461] (axis cs:11.7,0) rectangle (axis cs:12.3,3.08);
\draw[draw=none,fill=darkslategray532461] (axis cs:12.7,0) rectangle (axis cs:13.3,2.75);
\draw[draw=none,fill=darkslategray532461] (axis cs:13.7,0) rectangle (axis cs:14.3,2.51);
\draw[draw=none,fill=darkslategray532461] (axis cs:14.7,0) rectangle (axis cs:15.3,2.73);
\draw[draw=none,fill=darkslategray532461] (axis cs:15.7,0) rectangle (axis cs:16.3,2.42);
\draw[draw=none,fill=darkslategray532461] (axis cs:16.7,0) rectangle (axis cs:17.3,2.24);
\draw[draw=none,fill=darkslategray532461] (axis cs:17.7,0) rectangle (axis cs:18.3,2.37);
\draw[draw=none,fill=darkslategray532461] (axis cs:18.7,0) rectangle (axis cs:19.3,2.19);
\draw[draw=none,fill=darkslategray532461] (axis cs:19.7,0) rectangle (axis cs:20.3,1.99);
\draw[draw=none,fill=darkslategray532461] (axis cs:20.7,0) rectangle (axis cs:21.3,6.47);
\draw[draw=none,fill=darkslategray532461] (axis cs:21.7,0) rectangle (axis cs:22.3,7.2);
\draw[draw=none,fill=darkslategray532461] (axis cs:22.7,0) rectangle (axis cs:23.3,6.76);
\draw[draw=none,fill=darkslategray532461] (axis cs:23.7,0) rectangle (axis cs:24.3,6.41);
\draw[draw=none,fill=darkslategray532461] (axis cs:24.7,0) rectangle (axis cs:25.3,6.48);
\draw[draw=none,fill=darkslategray532461] (axis cs:25.7,0) rectangle (axis cs:26.3,6.18);
\draw[draw=none,fill=darkslategray532461] (axis cs:26.7,0) rectangle (axis cs:27.3,5.7);
\draw[draw=none,fill=darkslategray532461] (axis cs:27.7,0) rectangle (axis cs:28.3,5.81);
\draw[draw=none,fill=darkslategray532461] (axis cs:28.7,0) rectangle (axis cs:29.3,5.31);
\draw[draw=none,fill=darkslategray532461] (axis cs:29.7,0) rectangle (axis cs:30.3,4.95);
\draw[draw=none,fill=darkslategray532461] (axis cs:30.7,0) rectangle (axis cs:31.3,0);
\draw[draw=none,fill=darkslategray532461] (axis cs:31.7,0) rectangle (axis cs:32.3,0);
\draw[draw=none,fill=darkslategray532461] (axis cs:32.7,0) rectangle (axis cs:33.3,0);
\draw[draw=none,fill=darkslategray532461] (axis cs:33.7,0) rectangle (axis cs:34.3,0);
\draw[draw=none,fill=darkslategray532461] (axis cs:34.7,0) rectangle (axis cs:35.3,0.13);
\draw[draw=none,fill=firebrick1702141] (axis cs:0.7,0) rectangle (axis cs:1,0.03);
\addlegendimage{ybar,ybar legend,draw=none,fill=firebrick1702141}
\addlegendentry{2-stage brake, 1.5 s TTC}

\draw[draw=none,fill=firebrick1702141] (axis cs:1.7,0) rectangle (axis cs:2,0.07);
\draw[draw=none,fill=firebrick1702141] (axis cs:2.7,0) rectangle (axis cs:3,0.07);
\draw[draw=none,fill=firebrick1702141] (axis cs:3.7,0) rectangle (axis cs:4,0.03);
\draw[draw=none,fill=firebrick1702141] (axis cs:4.7,0) rectangle (axis cs:5,0.03);
\draw[draw=none,fill=firebrick1702141] (axis cs:5.7,0) rectangle (axis cs:6,0);
\draw[draw=none,fill=firebrick1702141] (axis cs:6.7,0) rectangle (axis cs:7,0);
\draw[draw=none,fill=firebrick1702141] (axis cs:7.7,0) rectangle (axis cs:8,0);
\draw[draw=none,fill=firebrick1702141] (axis cs:8.7,0) rectangle (axis cs:9,0);
\draw[draw=none,fill=firebrick1702141] (axis cs:9.7,0) rectangle (axis cs:10,0);
\draw[draw=none,fill=firebrick1702141] (axis cs:10.7,0) rectangle (axis cs:11,0);
\draw[draw=none,fill=firebrick1702141] (axis cs:11.7,0) rectangle (axis cs:12,0);
\draw[draw=none,fill=firebrick1702141] (axis cs:12.7,0) rectangle (axis cs:13,0);
\draw[draw=none,fill=firebrick1702141] (axis cs:13.7,0) rectangle (axis cs:14,0);
\draw[draw=none,fill=firebrick1702141] (axis cs:14.7,0) rectangle (axis cs:15,0);
\draw[draw=none,fill=firebrick1702141] (axis cs:15.7,0) rectangle (axis cs:16,0);
\draw[draw=none,fill=firebrick1702141] (axis cs:16.7,0) rectangle (axis cs:17,0);
\draw[draw=none,fill=firebrick1702141] (axis cs:17.7,0) rectangle (axis cs:18,0);
\draw[draw=none,fill=firebrick1702141] (axis cs:18.7,0) rectangle (axis cs:19,0);
\draw[draw=none,fill=firebrick1702141] (axis cs:19.7,0) rectangle (axis cs:20,0);
\draw[draw=none,fill=firebrick1702141] (axis cs:20.7,0) rectangle (axis cs:21,0.24);
\draw[draw=none,fill=firebrick1702141] (axis cs:21.7,0) rectangle (axis cs:22,0.22);
\draw[draw=none,fill=firebrick1702141] (axis cs:22.7,0) rectangle (axis cs:23,0.21);
\draw[draw=none,fill=firebrick1702141] (axis cs:23.7,0) rectangle (axis cs:24,0.2);
\draw[draw=none,fill=firebrick1702141] (axis cs:24.7,0) rectangle (axis cs:25,0.17);
\draw[draw=none,fill=firebrick1702141] (axis cs:25.7,0) rectangle (axis cs:26,0.16);
\draw[draw=none,fill=firebrick1702141] (axis cs:26.7,0) rectangle (axis cs:27,0.13);
\draw[draw=none,fill=firebrick1702141] (axis cs:27.7,0) rectangle (axis cs:28,0.13);
\draw[draw=none,fill=firebrick1702141] (axis cs:28.7,0) rectangle (axis cs:29,0.13);
\draw[draw=none,fill=firebrick1702141] (axis cs:29.7,0) rectangle (axis cs:30,0.07);
\draw[draw=none,fill=firebrick1702141] (axis cs:30.7,0) rectangle (axis cs:31,0);
\draw[draw=none,fill=firebrick1702141] (axis cs:31.7,0) rectangle (axis cs:32,0);
\draw[draw=none,fill=firebrick1702141] (axis cs:32.7,0) rectangle (axis cs:33,0);
\draw[draw=none,fill=firebrick1702141] (axis cs:33.7,0) rectangle (axis cs:34,0);
\draw[draw=none,fill=firebrick1702141] (axis cs:34.7,0) rectangle (axis cs:35,0);
\draw[draw=none,fill=coral24311881] (axis cs:1,0) rectangle (axis cs:1.3,0.5);
\addlegendimage{ybar,ybar legend,draw=none,fill=coral24311881}
\addlegendentry{2-stage brake, 1.25 s TTC}

\draw[draw=none,fill=coral24311881] (axis cs:2,0) rectangle (axis cs:2.3,0.6);
\draw[draw=none,fill=coral24311881] (axis cs:3,0) rectangle (axis cs:3.3,0.52);
\draw[draw=none,fill=coral24311881] (axis cs:4,0) rectangle (axis cs:4.3,0.48);
\draw[draw=none,fill=coral24311881] (axis cs:5,0) rectangle (axis cs:5.3,0.48);
\draw[draw=none,fill=coral24311881] (axis cs:6,0) rectangle (axis cs:6.3,0.46);
\draw[draw=none,fill=coral24311881] (axis cs:7,0) rectangle (axis cs:7.3,0.38);
\draw[draw=none,fill=coral24311881] (axis cs:8,0) rectangle (axis cs:8.3,0.44);
\draw[draw=none,fill=coral24311881] (axis cs:9,0) rectangle (axis cs:9.3,0.35);
\draw[draw=none,fill=coral24311881] (axis cs:10,0) rectangle (axis cs:10.3,0.33);
\draw[draw=none,fill=coral24311881] (axis cs:11,0) rectangle (axis cs:11.3,0.19);
\draw[draw=none,fill=coral24311881] (axis cs:12,0) rectangle (axis cs:12.3,0.37);
\draw[draw=none,fill=coral24311881] (axis cs:13,0) rectangle (axis cs:13.3,0.3);
\draw[draw=none,fill=coral24311881] (axis cs:14,0) rectangle (axis cs:14.3,0.25);
\draw[draw=none,fill=coral24311881] (axis cs:15,0) rectangle (axis cs:15.3,0.29);
\draw[draw=none,fill=coral24311881] (axis cs:16,0) rectangle (axis cs:16.3,0.24);
\draw[draw=none,fill=coral24311881] (axis cs:17,0) rectangle (axis cs:17.3,0.18);
\draw[draw=none,fill=coral24311881] (axis cs:18,0) rectangle (axis cs:18.3,0.18);
\draw[draw=none,fill=coral24311881] (axis cs:19,0) rectangle (axis cs:19.3,0.15);
\draw[draw=none,fill=coral24311881] (axis cs:20,0) rectangle (axis cs:20.3,0.15);
\draw[draw=none,fill=coral24311881] (axis cs:21,0) rectangle (axis cs:21.3,1.05);
\draw[draw=none,fill=coral24311881] (axis cs:22,0) rectangle (axis cs:22.3,1.17);
\draw[draw=none,fill=coral24311881] (axis cs:23,0) rectangle (axis cs:23.3,1.02);
\draw[draw=none,fill=coral24311881] (axis cs:24,0) rectangle (axis cs:24.3,0.97);
\draw[draw=none,fill=coral24311881] (axis cs:25,0) rectangle (axis cs:25.3,0.96);
\draw[draw=none,fill=coral24311881] (axis cs:26,0) rectangle (axis cs:26.3,0.91);
\draw[draw=none,fill=coral24311881] (axis cs:27,0) rectangle (axis cs:27.3,0.84);
\draw[draw=none,fill=coral24311881] (axis cs:28,0) rectangle (axis cs:28.3,0.8);
\draw[draw=none,fill=coral24311881] (axis cs:29,0) rectangle (axis cs:29.3,0.7);
\draw[draw=none,fill=coral24311881] (axis cs:30,0) rectangle (axis cs:30.3,0.67);
\draw[draw=none,fill=coral24311881] (axis cs:31,0) rectangle (axis cs:31.3,0);
\draw[draw=none,fill=coral24311881] (axis cs:32,0) rectangle (axis cs:32.3,0);
\draw[draw=none,fill=coral24311881] (axis cs:33,0) rectangle (axis cs:33.3,0);
\draw[draw=none,fill=coral24311881] (axis cs:34,0) rectangle (axis cs:34.3,0);
\draw[draw=none,fill=coral24311881] (axis cs:35,0) rectangle (axis cs:35.3,0);

\nextgroupplot[
tick align=outside,
tick pos=left,
x grid style={darkgray176},
xlabel={Scenario number},
xmajorgrids,
xmin=-1.03, xmax=37.03,
xtick style={color=black},
y grid style={darkgray176},
ylabel={Sev. opp [\%]},
ymajorgrids,
ymin=0, ymax=13,
ytick style={color=black},
height = 3.5cm,
width = \textwidth,
font=\scriptsize
]
\draw[draw=none,fill=darkslategray532461] (axis cs:0.7,0) rectangle (axis cs:1.3,4.55);
\draw[draw=none,fill=darkslategray532461] (axis cs:1.7,0) rectangle (axis cs:2.3,5.19);
\draw[draw=none,fill=darkslategray532461] (axis cs:2.7,0) rectangle (axis cs:3.3,4.93);
\draw[draw=none,fill=darkslategray532461] (axis cs:3.7,0) rectangle (axis cs:4.3,4.53);
\draw[draw=none,fill=darkslategray532461] (axis cs:4.7,0) rectangle (axis cs:5.3,4.6);
\draw[draw=none,fill=darkslategray532461] (axis cs:5.7,0) rectangle (axis cs:6.3,4.23);
\draw[draw=none,fill=darkslategray532461] (axis cs:6.7,0) rectangle (axis cs:7.3,3.92);
\draw[draw=none,fill=darkslategray532461] (axis cs:7.7,0) rectangle (axis cs:8.3,3.98);
\draw[draw=none,fill=darkslategray532461] (axis cs:8.7,0) rectangle (axis cs:9.3,3.68);
\draw[draw=none,fill=darkslategray532461] (axis cs:9.7,0) rectangle (axis cs:10.3,3.45);
\draw[draw=none,fill=darkslategray532461] (axis cs:10.7,0) rectangle (axis cs:11.3,2.95);
\draw[draw=none,fill=darkslategray532461] (axis cs:11.7,0) rectangle (axis cs:12.3,3.87);
\draw[draw=none,fill=darkslategray532461] (axis cs:12.7,0) rectangle (axis cs:13.3,3.45);
\draw[draw=none,fill=darkslategray532461] (axis cs:13.7,0) rectangle (axis cs:14.3,3.25);
\draw[draw=none,fill=darkslategray532461] (axis cs:14.7,0) rectangle (axis cs:15.3,3.56);
\draw[draw=none,fill=darkslategray532461] (axis cs:15.7,0) rectangle (axis cs:16.3,3.17);
\draw[draw=none,fill=darkslategray532461] (axis cs:16.7,0) rectangle (axis cs:17.3,2.91);
\draw[draw=none,fill=darkslategray532461] (axis cs:17.7,0) rectangle (axis cs:18.3,3.1);
\draw[draw=none,fill=darkslategray532461] (axis cs:18.7,0) rectangle (axis cs:19.3,2.88);
\draw[draw=none,fill=darkslategray532461] (axis cs:19.7,0) rectangle (axis cs:20.3,2.63);
\draw[draw=none,fill=darkslategray532461] (axis cs:20.7,0) rectangle (axis cs:21.3,7.11);
\draw[draw=none,fill=darkslategray532461] (axis cs:21.7,0) rectangle (axis cs:22.3,7.83);
\draw[draw=none,fill=darkslategray532461] (axis cs:22.7,0) rectangle (axis cs:23.3,7.42);
\draw[draw=none,fill=darkslategray532461] (axis cs:23.7,0) rectangle (axis cs:24.3,7.11);
\draw[draw=none,fill=darkslategray532461] (axis cs:24.7,0) rectangle (axis cs:25.3,7.19);
\draw[draw=none,fill=darkslategray532461] (axis cs:25.7,0) rectangle (axis cs:26.3,7.02);
\draw[draw=none,fill=darkslategray532461] (axis cs:26.7,0) rectangle (axis cs:27.3,6.55);
\draw[draw=none,fill=darkslategray532461] (axis cs:27.7,0) rectangle (axis cs:28.3,6.57);
\draw[draw=none,fill=darkslategray532461] (axis cs:28.7,0) rectangle (axis cs:29.3,6.11);
\draw[draw=none,fill=darkslategray532461] (axis cs:29.7,0) rectangle (axis cs:30.3,5.74);
\draw[draw=none,fill=darkslategray532461] (axis cs:30.7,0) rectangle (axis cs:31.3,5.92);
\draw[draw=none,fill=darkslategray532461] (axis cs:31.7,0) rectangle (axis cs:32.3,8.49);
\draw[draw=none,fill=darkslategray532461] (axis cs:32.7,0) rectangle (axis cs:33.3,0.51);
\draw[draw=none,fill=darkslategray532461] (axis cs:33.7,0) rectangle (axis cs:34.3,12.75);
\draw[draw=none,fill=darkslategray532461] (axis cs:34.7,0) rectangle (axis cs:35.3,0.24);
\draw[draw=none,fill=firebrick1702141] (axis cs:0.7,0) rectangle (axis cs:1,0.06);
\draw[draw=none,fill=firebrick1702141] (axis cs:1.7,0) rectangle (axis cs:2,0.12);
\draw[draw=none,fill=firebrick1702141] (axis cs:2.7,0) rectangle (axis cs:3,0.12);
\draw[draw=none,fill=firebrick1702141] (axis cs:3.7,0) rectangle (axis cs:4,0.06);
\draw[draw=none,fill=firebrick1702141] (axis cs:4.7,0) rectangle (axis cs:5,0.06);
\draw[draw=none,fill=firebrick1702141] (axis cs:5.7,0) rectangle (axis cs:6,0);
\draw[draw=none,fill=firebrick1702141] (axis cs:6.7,0) rectangle (axis cs:7,0);
\draw[draw=none,fill=firebrick1702141] (axis cs:7.7,0) rectangle (axis cs:8,0);
\draw[draw=none,fill=firebrick1702141] (axis cs:8.7,0) rectangle (axis cs:9,0);
\draw[draw=none,fill=firebrick1702141] (axis cs:9.7,0) rectangle (axis cs:10,0);
\draw[draw=none,fill=firebrick1702141] (axis cs:10.7,0) rectangle (axis cs:11,0);
\draw[draw=none,fill=firebrick1702141] (axis cs:11.7,0) rectangle (axis cs:12,0);
\draw[draw=none,fill=firebrick1702141] (axis cs:12.7,0) rectangle (axis cs:13,0);
\draw[draw=none,fill=firebrick1702141] (axis cs:13.7,0) rectangle (axis cs:14,0);
\draw[draw=none,fill=firebrick1702141] (axis cs:14.7,0) rectangle (axis cs:15,0);
\draw[draw=none,fill=firebrick1702141] (axis cs:15.7,0) rectangle (axis cs:16,0);
\draw[draw=none,fill=firebrick1702141] (axis cs:16.7,0) rectangle (axis cs:17,0);
\draw[draw=none,fill=firebrick1702141] (axis cs:17.7,0) rectangle (axis cs:18,0);
\draw[draw=none,fill=firebrick1702141] (axis cs:18.7,0) rectangle (axis cs:19,0);
\draw[draw=none,fill=firebrick1702141] (axis cs:19.7,0) rectangle (axis cs:20,0);
\draw[draw=none,fill=firebrick1702141] (axis cs:20.7,0) rectangle (axis cs:21,0.4);
\draw[draw=none,fill=firebrick1702141] (axis cs:21.7,0) rectangle (axis cs:22,0.33);
\draw[draw=none,fill=firebrick1702141] (axis cs:22.7,0) rectangle (axis cs:23,0.33);
\draw[draw=none,fill=firebrick1702141] (axis cs:23.7,0) rectangle (axis cs:24,0.32);
\draw[draw=none,fill=firebrick1702141] (axis cs:24.7,0) rectangle (axis cs:25,0.26);
\draw[draw=none,fill=firebrick1702141] (axis cs:25.7,0) rectangle (axis cs:26,0.25);
\draw[draw=none,fill=firebrick1702141] (axis cs:26.7,0) rectangle (axis cs:27,0.2);
\draw[draw=none,fill=firebrick1702141] (axis cs:27.7,0) rectangle (axis cs:28,0.2);
\draw[draw=none,fill=firebrick1702141] (axis cs:28.7,0) rectangle (axis cs:29,0.2);
\draw[draw=none,fill=firebrick1702141] (axis cs:29.7,0) rectangle (axis cs:30,0.13);
\draw[draw=none,fill=firebrick1702141] (axis cs:30.7,0) rectangle (axis cs:31,0);
\draw[draw=none,fill=firebrick1702141] (axis cs:31.7,0) rectangle (axis cs:32,0);
\draw[draw=none,fill=firebrick1702141] (axis cs:32.7,0) rectangle (axis cs:33,0);
\draw[draw=none,fill=firebrick1702141] (axis cs:33.7,0) rectangle (axis cs:34,0);
\draw[draw=none,fill=firebrick1702141] (axis cs:34.7,0) rectangle (axis cs:35,0);
\draw[draw=none,fill=coral24311881] (axis cs:1,0) rectangle (axis cs:1.3,0.74);
\draw[draw=none,fill=coral24311881] (axis cs:2,0) rectangle (axis cs:2.3,0.91);
\draw[draw=none,fill=coral24311881] (axis cs:3,0) rectangle (axis cs:3.3,0.78);
\draw[draw=none,fill=coral24311881] (axis cs:4,0) rectangle (axis cs:4.3,0.68);
\draw[draw=none,fill=coral24311881] (axis cs:5,0) rectangle (axis cs:5.3,0.71);
\draw[draw=none,fill=coral24311881] (axis cs:6,0) rectangle (axis cs:6.3,0.66);
\draw[draw=none,fill=coral24311881] (axis cs:7,0) rectangle (axis cs:7.3,0.57);
\draw[draw=none,fill=coral24311881] (axis cs:8,0) rectangle (axis cs:8.3,0.64);
\draw[draw=none,fill=coral24311881] (axis cs:9,0) rectangle (axis cs:9.3,0.49);
\draw[draw=none,fill=coral24311881] (axis cs:10,0) rectangle (axis cs:10.3,0.48);
\draw[draw=none,fill=coral24311881] (axis cs:11,0) rectangle (axis cs:11.3,0.27);
\draw[draw=none,fill=coral24311881] (axis cs:12,0) rectangle (axis cs:12.3,0.5);
\draw[draw=none,fill=coral24311881] (axis cs:13,0) rectangle (axis cs:13.3,0.42);
\draw[draw=none,fill=coral24311881] (axis cs:14,0) rectangle (axis cs:14.3,0.35);
\draw[draw=none,fill=coral24311881] (axis cs:15,0) rectangle (axis cs:15.3,0.41);
\draw[draw=none,fill=coral24311881] (axis cs:16,0) rectangle (axis cs:16.3,0.34);
\draw[draw=none,fill=coral24311881] (axis cs:17,0) rectangle (axis cs:17.3,0.27);
\draw[draw=none,fill=coral24311881] (axis cs:18,0) rectangle (axis cs:18.3,0.27);
\draw[draw=none,fill=coral24311881] (axis cs:19,0) rectangle (axis cs:19.3,0.21);
\draw[draw=none,fill=coral24311881] (axis cs:20,0) rectangle (axis cs:20.3,0.21);
\draw[draw=none,fill=coral24311881] (axis cs:21,0) rectangle (axis cs:21.3,1.59);
\draw[draw=none,fill=coral24311881] (axis cs:22,0) rectangle (axis cs:22.3,1.69);
\draw[draw=none,fill=coral24311881] (axis cs:23,0) rectangle (axis cs:23.3,1.51);
\draw[draw=none,fill=coral24311881] (axis cs:24,0) rectangle (axis cs:24.3,1.43);
\draw[draw=none,fill=coral24311881] (axis cs:25,0) rectangle (axis cs:25.3,1.42);
\draw[draw=none,fill=coral24311881] (axis cs:26,0) rectangle (axis cs:26.3,1.35);
\draw[draw=none,fill=coral24311881] (axis cs:27,0) rectangle (axis cs:27.3,1.22);
\draw[draw=none,fill=coral24311881] (axis cs:28,0) rectangle (axis cs:28.3,1.19);
\draw[draw=none,fill=coral24311881] (axis cs:29,0) rectangle (axis cs:29.3,1.01);
\draw[draw=none,fill=coral24311881] (axis cs:30,0) rectangle (axis cs:30.3,0.99);
\draw[draw=none,fill=coral24311881] (axis cs:31,0) rectangle (axis cs:31.3,0.8);
\draw[draw=none,fill=coral24311881] (axis cs:32,0) rectangle (axis cs:32.3,1.2);
\draw[draw=none,fill=coral24311881] (axis cs:33,0) rectangle (axis cs:33.3,0);
\draw[draw=none,fill=coral24311881] (axis cs:34,0) rectangle (axis cs:34.3,1.89);
\draw[draw=none,fill=coral24311881] (axis cs:35,0) rectangle (axis cs:35.3,0);
\end{groupplot}

\end{tikzpicture}